\documentclass{sig-alternate-2013}
\usepackage{subfig, float, graphicx, epstopdf, balance}
\usepackage{color, soul}
\usepackage[table]{xcolor}
\usepackage{epsfig}
\usepackage{booktabs}

\usepackage{mathtools, url}
\usepackage{latexsym}
\usepackage{amsfonts} 
\usepackage{adjustbox}
\usepackage{lipsum}

\def\P{\mathbb{P}}

\usepackage{float} 
\usepackage{amsmath}
\usepackage{amssymb}
\usepackage{amsfonts}

\usepackage{multirow}

\usepackage[table]{xcolor}
\usepackage{epsfig}
\usepackage{booktabs, balance}
\usepackage{adjustbox, url}

\def\P{\mathbb{P}}

\definecolor{lightgray}{gray}{0.85}

\begin{document}

\title{Travel the World: Analyzing and Predicting \\Booking Behavior using E-mail Travel Receipts}


\author{
\alignauthor Nemanja Djuric$^\dag$, Mihajlo Grbovic$^\dag$, Vladan Radosavljevic$^\dag$, \\Jaikit Savla$^\ddag$, Varun Bhagwan$^\ddag$, Doug Sharp$^\ddag$ \\
\affaddr{$^\dag$Yahoo Labs, Sunnyvale, CA, USA ~~~~~~~$^\ddag$Yahoo Inc., Sunnyvale, CA, USA}\\
\email{\{nemanja, mihajlo, vladan, jaikit, vbhagwan, dsharp\}@yahoo-inc.com}
}



\permission{Copyright is held by the author/owner(s).}
\conferenceinfo{WWW'16 Companion,}{April 11--15, 2016, Montr\'eal, Qu\'ebec, Canada.} 
\copyrightetc{ACM \the\acmcopyr}
\crdata{978-1-4503-4144-8/16/04. \\
http://dx.doi.org/10.1145/2872518.2889410}

\clubpenalty=10000 
\widowpenalty = 10000

\maketitle
\begin{abstract}
Tourism industry has grown tremendously in the previous several decades. Despite its global impact, there still remain a number of open questions related to better understanding of tourists and their habits. In this work we analyze the largest data set of travel receipts considered thus far, and focus on exploring and modeling booking behavior of online customers. We extract useful, actionable insights into the booking behavior, and tackle the task of predicting the booking time. The presented results can be directly used to improve booking experience of customers and optimize targeting campaigns of travel operators. 
\end{abstract}


\section{Introduction}
Tourism and hospitality sectors have witnessed enormous growth in the past century. Originating as a pastime of the wealthy, only very recently technological advances and global economic progress have brought traveling for leisure closer to nearly all layers of the society \cite{sezgin2012golden}. The rise from modest beginnings has been astounding, and today tourism accounts for nearly $10\%$ of the world's gross domestic product, employing more than $277$ million people worldwide \cite{wttc2015}. The upward trend is projected to continue in the future at unabated rates, further highlighting the impact and the importance of the thriving industry.

Given such deep economic footprint, it is highly beneficial to understand customers' behavior in order to satisfy the growing demand. A number of challenging problems has brought attention of researchers, who proposed diverse approaches towards painting a clearer picture of a modern tourist. More specifically, in \cite{oppermann1995model} authors focus on understanding how tourists make their travel plans, while authors of \cite{min2002data} tackle the task of profiling of hotel patrons and learning actionable behavioral patterns using rule mining. In the absence of travel data, some studies attempt to recreate travel information using online photos and other resources \cite{zheng2012mining}.

In addition to modeling tourist behavior when they are already on the road, understanding decision process that led to booking of the trip is equally important, as it allows tourist companies to actively influence decisions of a potential customer at its source through better ad targeting \cite{rudstrom2003socially}. For example, effect of online reviews on booking decisions is explored in \cite{sparks2011impact}, while \cite{matzler2005consequences} analyzes impact of information overload. However, such studies mostly rely on surveys comprising few hundred examples, limiting their generality. We address this issue, and explore booking behavior of millions of online customers by mining their travel receipts.

\begin{figure}[t!]
  \centering
  \subfloat{\label{fig:weekly_travel}\includegraphics[width=0.20\textwidth]{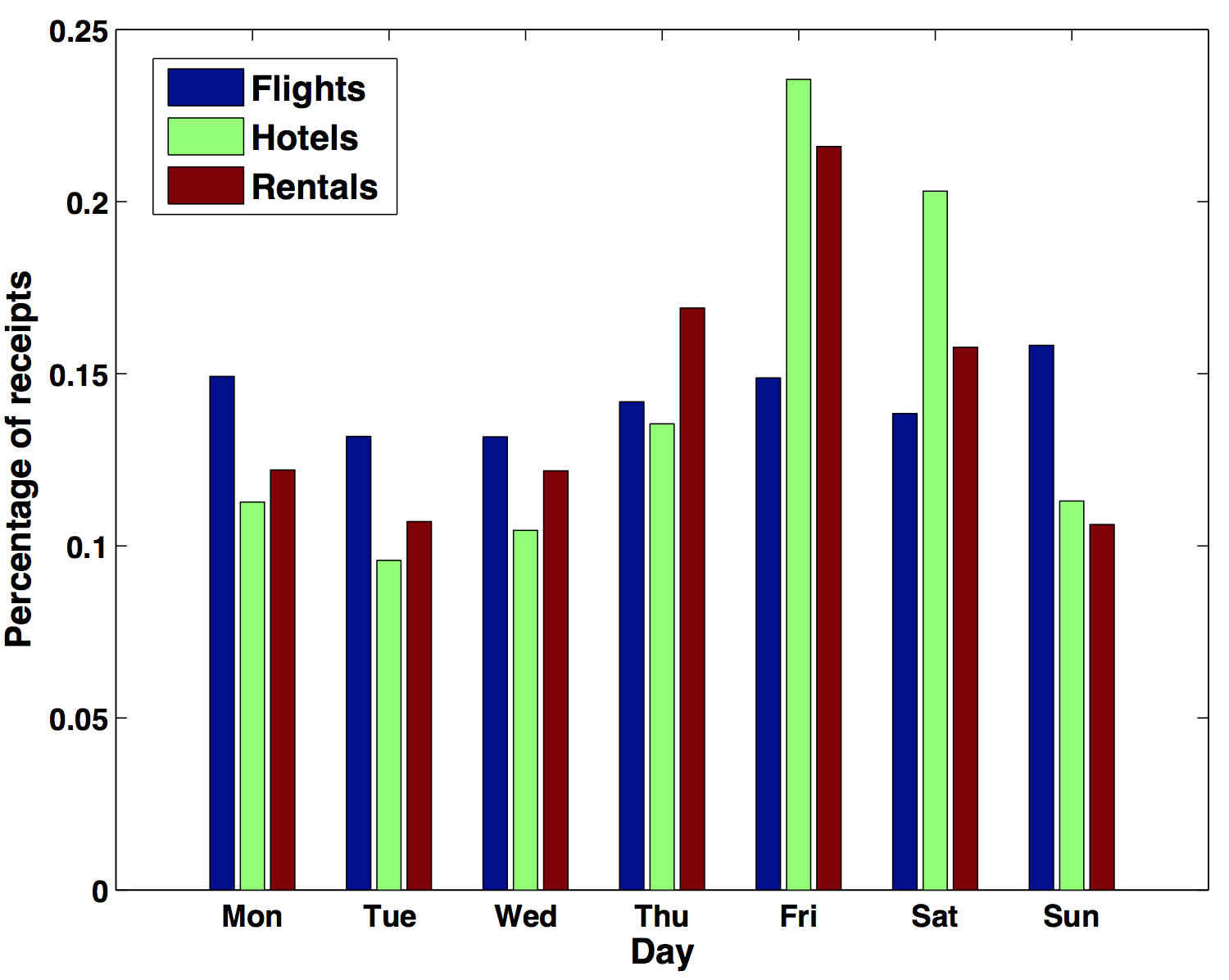}}~~
  \subfloat{\label{fig:weekly_booking}\includegraphics[width=0.20\textwidth]{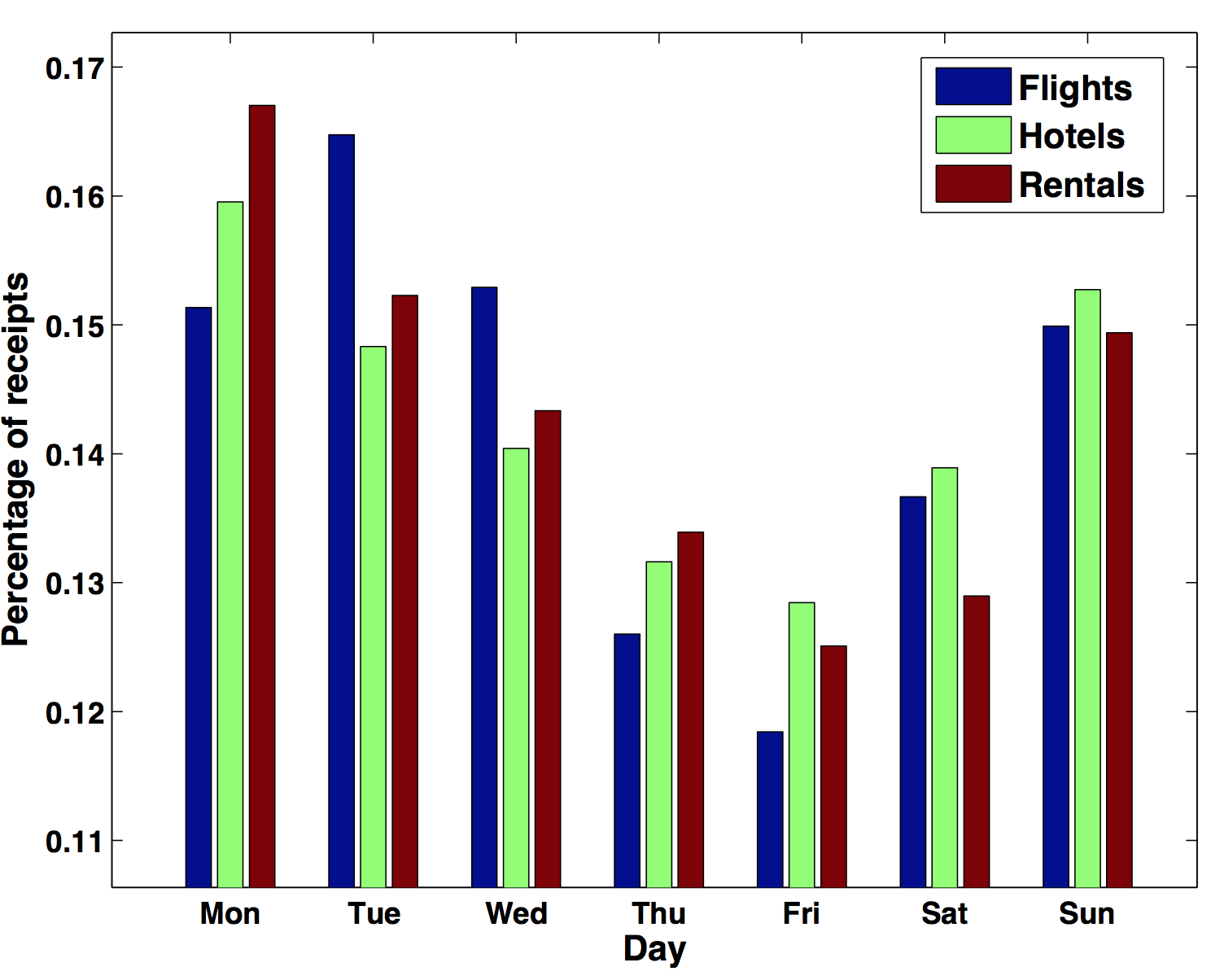}}
  \vspace{-2mm}
  \caption{Distribution of a) check-in; and b) booking days}
  \label{fig:weekly_dist}
\end{figure}

\section{Analysis of booking behavior}
We collected more than 25 million travel receipts, received by a subset of Yahoo Mail users from January through April, 2015. We considered flight, hotel, and car rental receipts, for which we extracted booking timestamp, travel (for flights) and check-in and check-out dates (for hotels and rentals). The data was anonymized (users were assigned random IDs).

\begin{figure*}[t!]
  \centering
  \subfloat{\label{fig:cumul_flight}\includegraphics[width=0.22\textwidth]{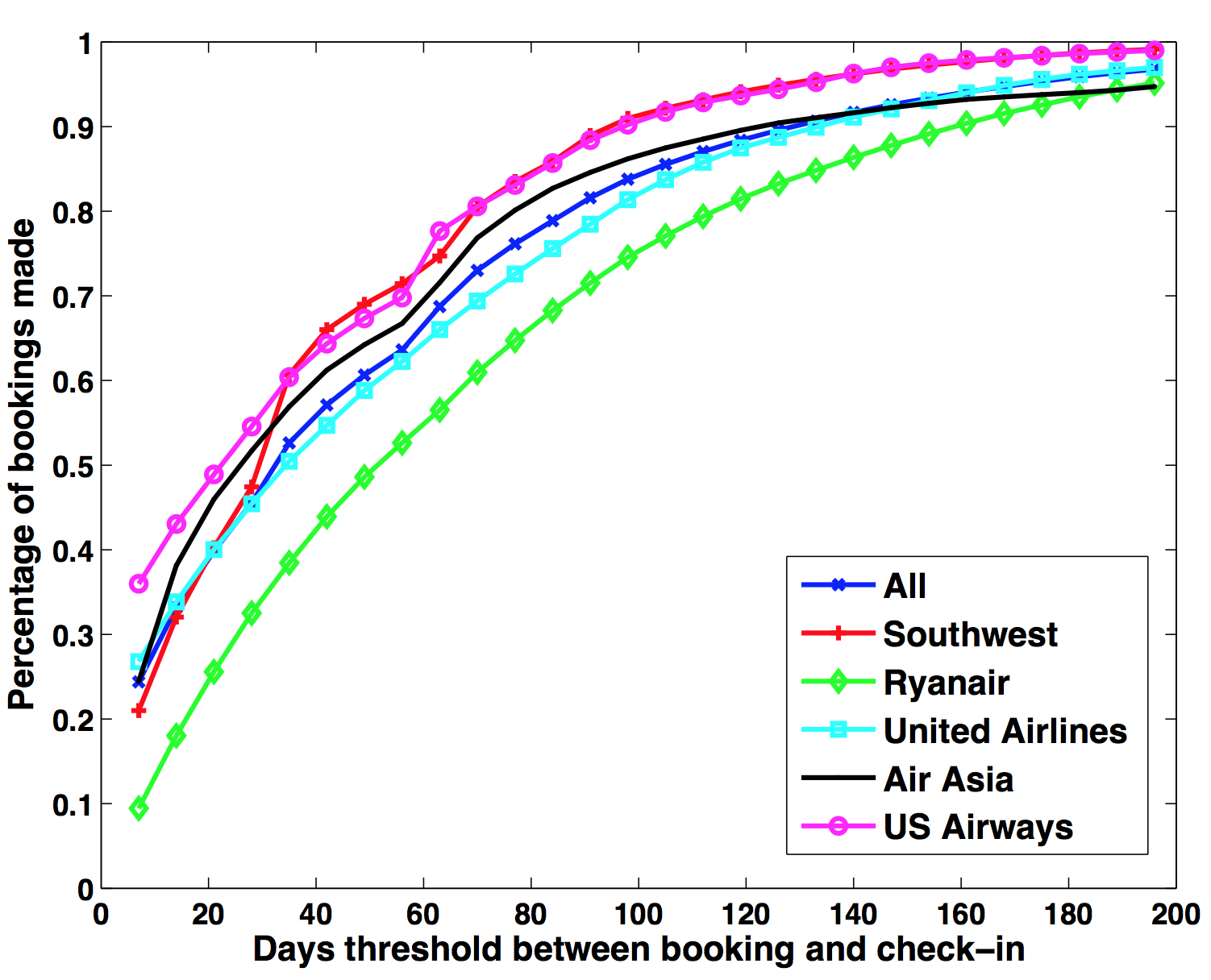}}~~~
  \subfloat{\label{fig:cumul_hotel}\includegraphics[width=0.22\textwidth]{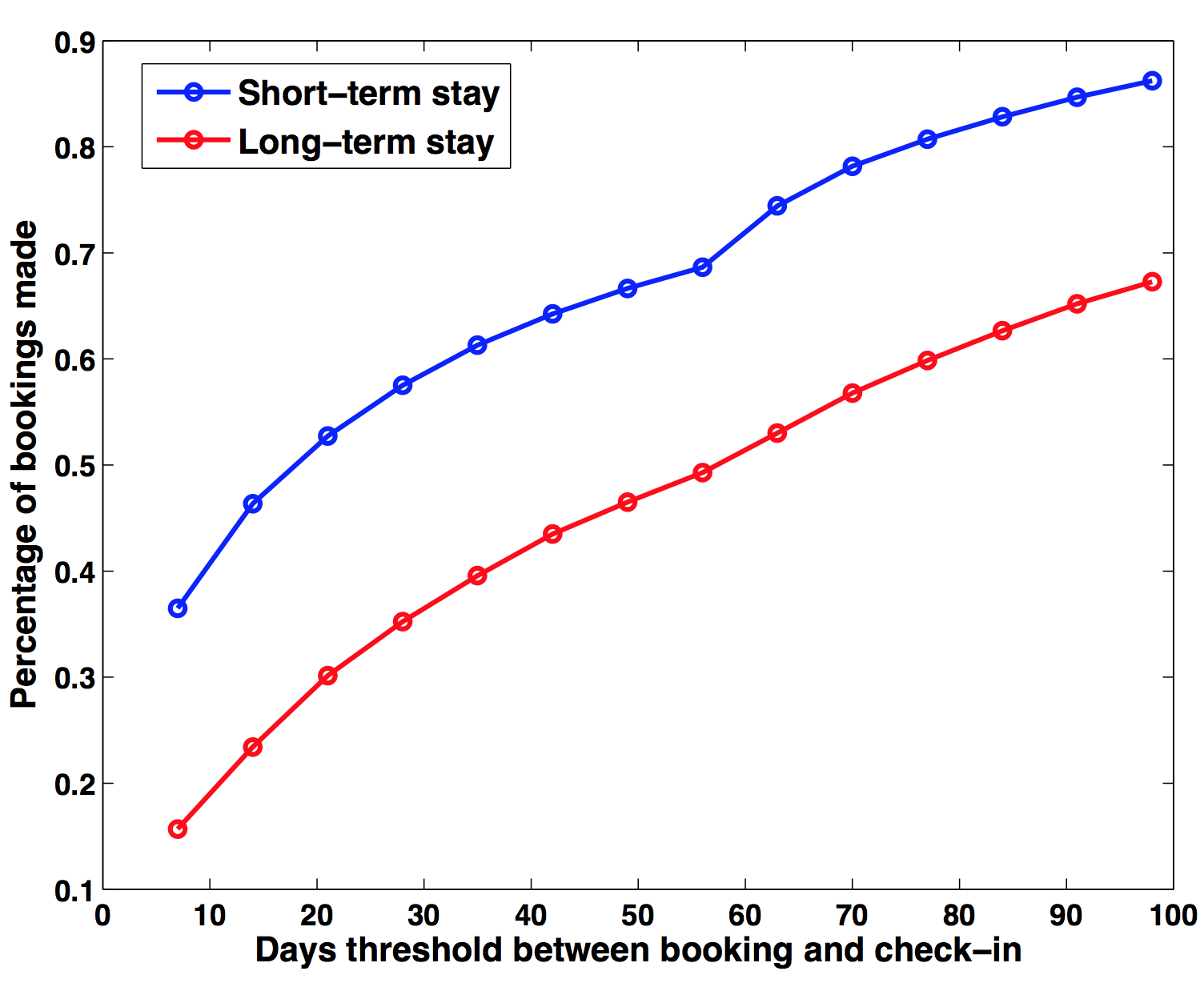}}~~~
  \subfloat{\label{fig:cumul_rental}\includegraphics[width=0.22\textwidth]{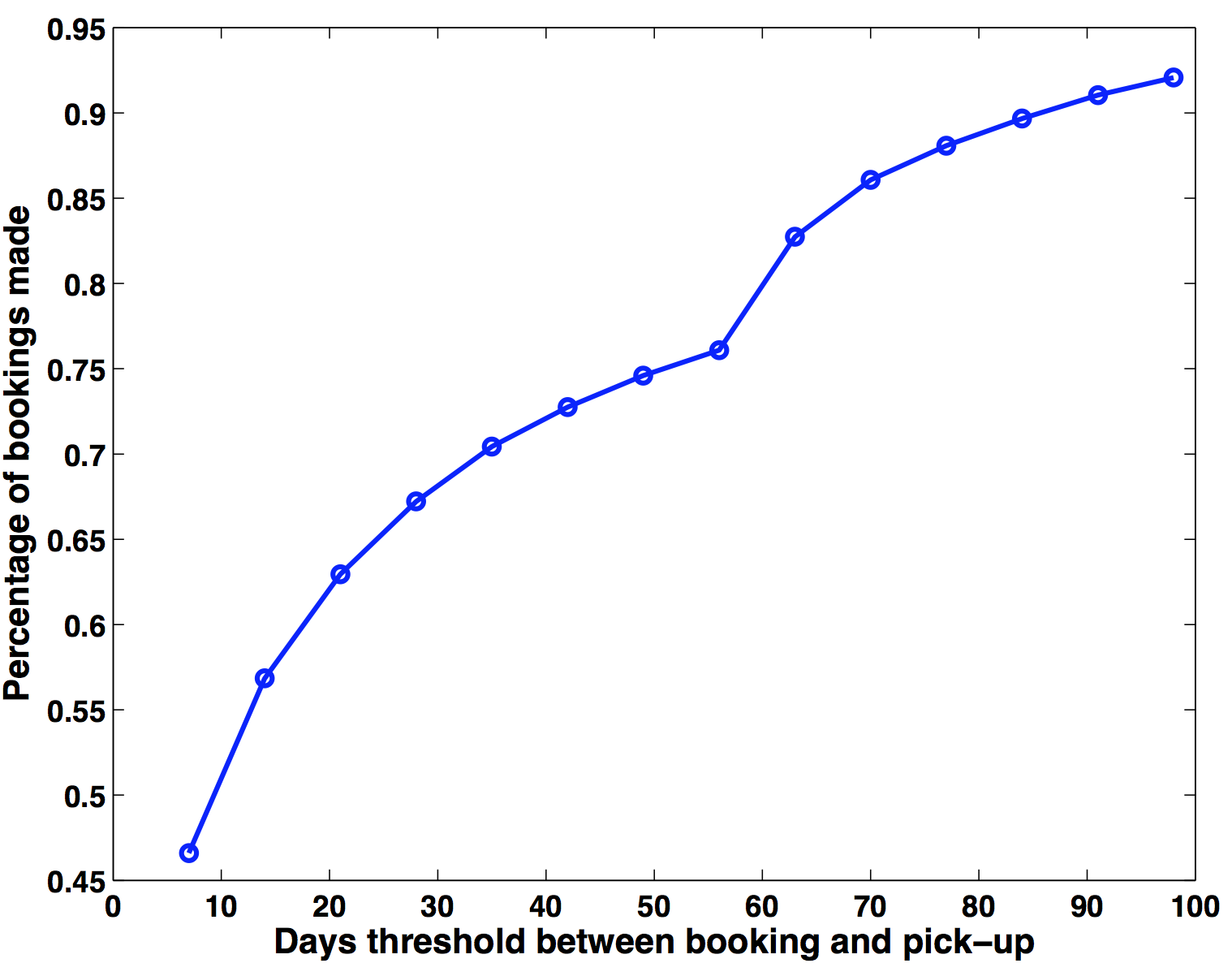}}
	\vspace{-3mm}
  \caption{Cumulative distribution of receipts with respect to days booked in advance: a) flights; b) hotels; c) rentals}
  \label{fig:cumul_receipt}
	\vspace{-6mm}
\end{figure*}

\begin{figure*}[t!]
  \centering
  \subfloat[Hotel vs. flight]{\label{fig:diff_haf}\includegraphics[width=0.22\textwidth]{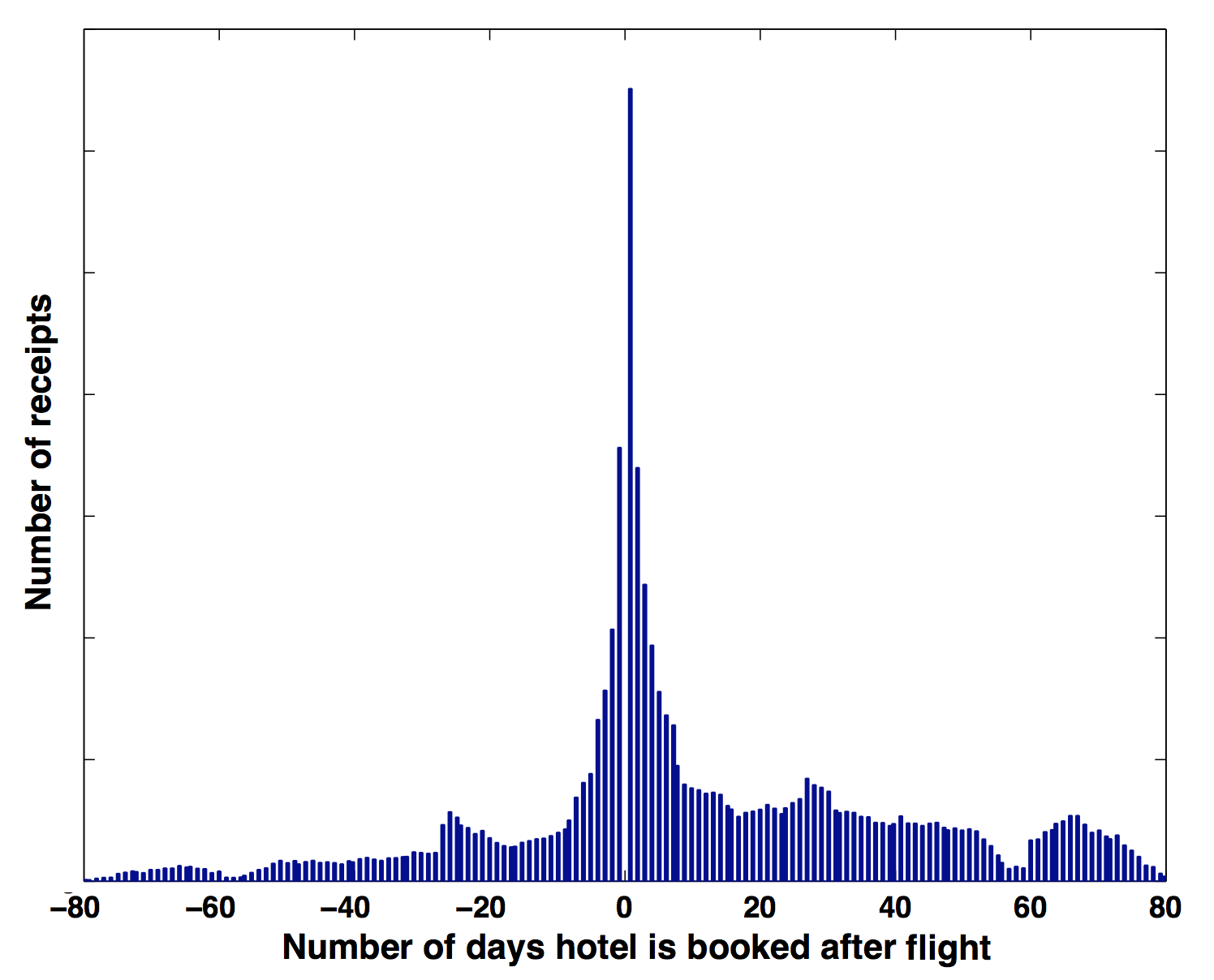}}
  \subfloat[Rental vs. hotel]{\label{fig:diff_har}~~~~~\includegraphics[width=0.22\textwidth]{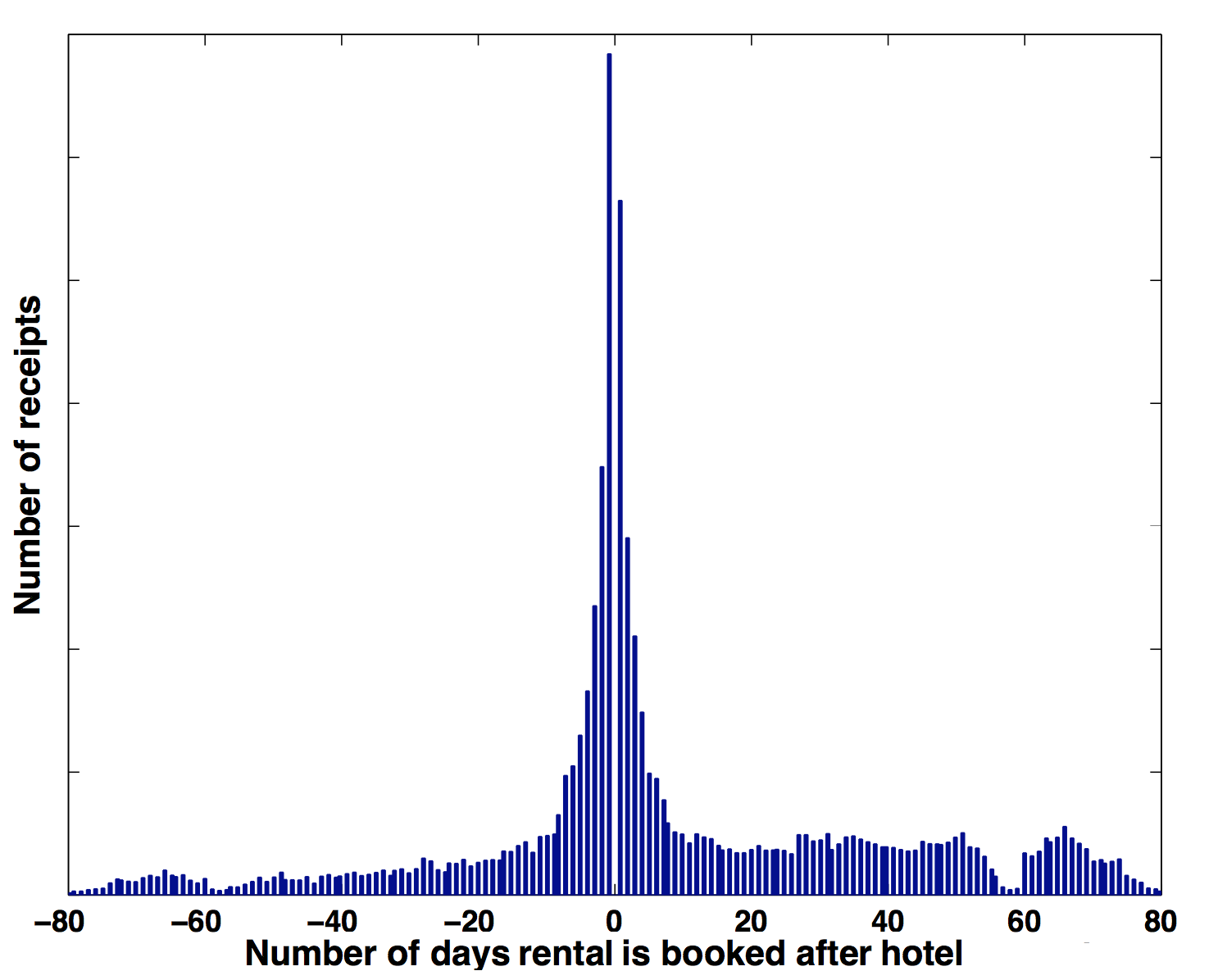}}
  \subfloat[Rental vs. flight]{\label{fig:diff_raf}~~~~~\includegraphics[width=0.215\textwidth]{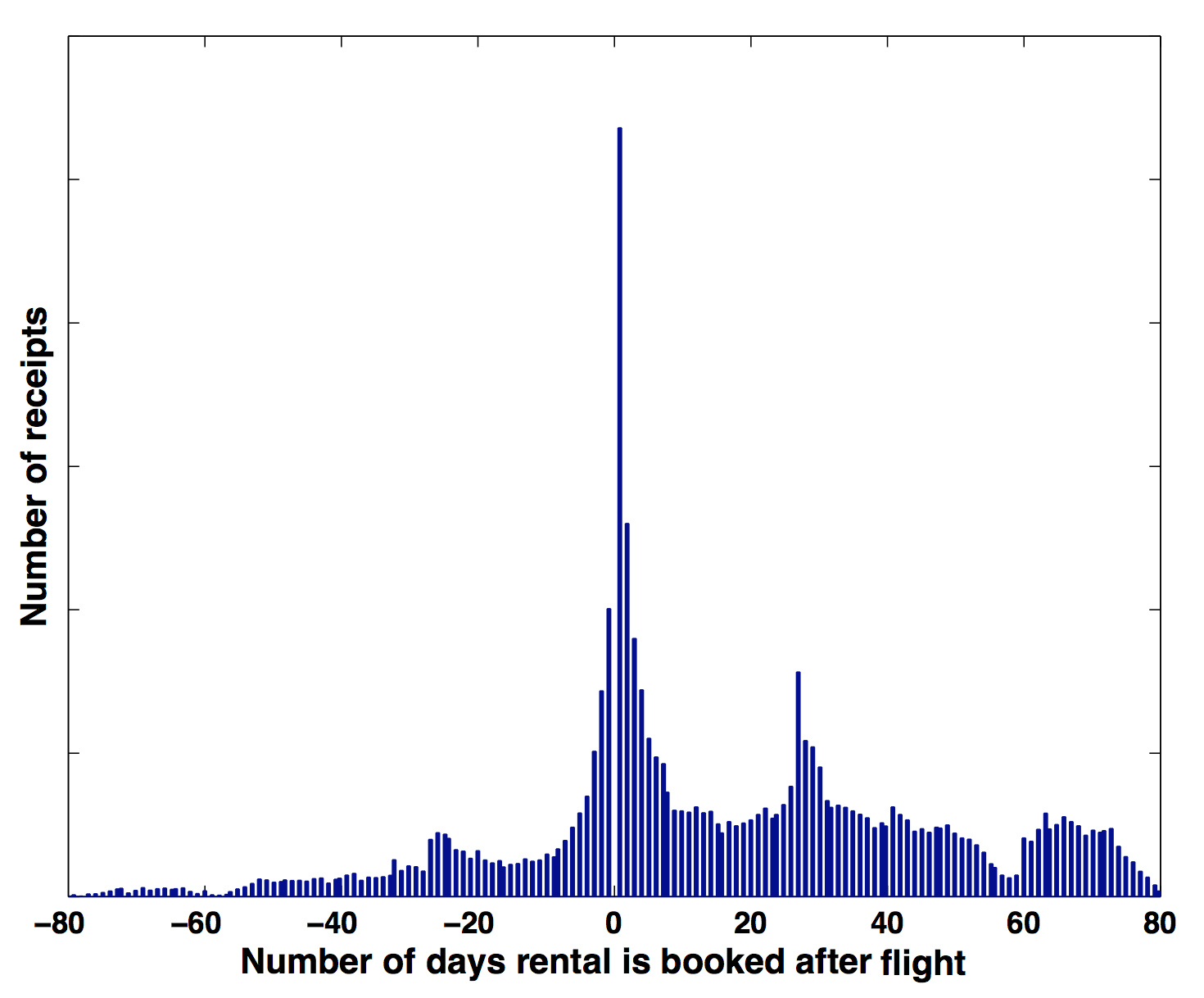}}
	\vspace{-3mm}
  \caption{Number of days between coupled bookings from different channels}
  \label{fig:diffs}
	\vspace{-4mm}
\end{figure*}

First, let us consider weekly distribution of check-in and booking days. In Figure \ref{fig:weekly_travel} we can see that the largest number of hotel check-ins and rental pick-ups happens on Fridays and Saturdays, while flights had largely uniform distribution. On the other hand, as seen in Figure \ref{fig:weekly_booking}, the largest number of bookings for hotels and rentals is made on Mondays, and on Tuesdays for flights. The least number of bookings in all cases is made on Fridays.
Another interesting finding was a huge increase in hotel check-ins on Valentine's day (not shown), more than 2 times higher number than rates one week before or after. Curiously, we did not see the corresponding increase in flights and rentals, implying that most of these hotel stays were local.

  \vspace{-1mm}
\subsection{Analysis of individual channels}
We take a closer look at the booking channels independently. In Figure \ref{fig:cumul_flight} we analyze airline booking times segregated by companies. Around 22\% of flights are booked within one week ahead of the trip. However, when we take a look at each company individually, this fluctuates from 10\% for RyanAir to 35\% for US Airways. The results suggests that customers plan further ahead for lower-cost companies which often have special deals and promotions in such cases.

For hotels, we can see that there is a significant difference between short-term (less than 7 days) and long-term stay (7 days and more) in terms of number of days the rooms are booked in advance. For short-term stay nearly 40\% of all bookings are made within one week in advance, while for long-term stay the percentage drops to 15\%. In other words, customers are planning more in advance for longer, more expensive stays. We also observe jumps in the number of bookings 60 days in advance, which could be explained by black-out windows imposed by some loyalty programs.

Lastly, we see that more than 45\% of rentals are booked within 7 days from car pick-up date. This grows to 70\% one month prior to pick-up, indicating that rentals are planned much less ahead than either flights or hotels.

\subsection{Analysis of cross-channel correlation}
Next we analyzed time difference between bookings from different channels made for the same trip, where the bookings were deemed part of the same trip if the check-in times are within 24 hours. We only considered trips where the difference between booking dates and check-in date is longer than 30 days, in order to avoid the effect of late purchases which would bias the bookings to be temporally close. The results are given in Figure \ref{fig:diffs}, where $x$-axis shows number of days between bookings from two different channels. 

By considering the three subfigures, we can see that the majority of two-channel bookings are made within 10 days from each other. Moreover, we can conclude that when users are booking more than one channel for a single trip (e.g., booking a flight in addition to hotel, or rental in addition to flight), they are mostly purchasing them in the order ``flight $\rightarrow$ hotel $\rightarrow$ rental", going from higher demand toward smaller demand channels. Interestingly, this conclusion is also confirmed by Figure \ref{fig:cumul_receipt}, where 50\% of flights are booked around 30 days in advance, 50\% of short-term hotel stays around 20 days in advance, while half of car rentals are booked 10 days in advance. These insights are of great importance to booking agencies (e.g., they provide guidance on how and when to target users that already booked a flight).

\section{Predicting booking behavior}
In this section we turn our attention to behavior prediction task. In particular, we predict when a user will book a rental given they already booked a flight. We cast the problem as a classification task and predict labels ``<5 days", ``6-10", ``11-20", ``21-40", and ``>41 or never", while we used age, gender, and past bookings in each channel as features. We trained logistic regression (we split the data into equally-sized training and test sets), and compared to baselines that predict most frequent label and most frequent label in age-gender groups (age buckets were set to ``<26", ``26-40", ``>40"). The results are given in Table \ref{tbl:classification}, where we see that the proposed approach outperformed the baselines.

\begin{table}[t]
\caption{Predicting rental time given flight booking}
\vspace{-6mm}
\begin{center}
\begin{tabular}{l c}
~~~~~Method & Rel. accuracy improv. \\
\hline \hline
\rowcolor{lightgray}
Most frequent & 0\%  \\
Age-gender bucket & 14\% \\
\rowcolor{lightgray}
Logistic regression & 23\%  \\
\bottomrule
\end{tabular}
\end{center}
\label{tbl:classification}
\end{table}

%
\bibliographystyle{abbrv}
{
\bibliography{references}  
}

\end{document}